\documentclass[11pt]{article}

\usepackage{fullpage,calc,amsfonts,amssymb,amsmath,bm,url,color,theorem}
\usepackage{graphicx,subfigure}
\usepackage{epsfig,multirow}
\usepackage{algorithmic,algorithm}
\usepackage{setspace}

{\begin{list}{}{
    \settowidth{\labelwidth}{\mbox{\textnormal{#1}}}%
    \setlength{\leftmargin}{\labelwidth+\labelsep}}}%
{\end{list}}

    \makeatletter
    \def\multilimits@{\bgroup
  \Let@
  \restore@math@cr
  \default@tag
 \baselineskip\fontdimen10 \scriptfont\tw@
 \advance\baselineskip\fontdimen12 \scriptfont\tw@
 \lineskip\thr@@\fontdimen8 \scriptfont\thr@@
 \lineskiplimit\lineskip
 \vbox\bgroup\ialign\bgroup\hfil$\m@th\scriptstyle{##}$\hfil\crcr}
    \def\Sb{_\multilimits@}
    \def\endSb{\crcr\egroup\egroup\egroup}
\makeatother

\newtheorem{Prop}{Proposition}

\theorembodyfont{\rmfamily}

\begin{document}

\bibliographystyle{IEEEtran}

%-------------
\vspace*{0.25in}

\begin{center}
\doublespacing
{\LARGE
Achievable Rate Derivations and Further Simulation Results for
``Physical-Layer Multicasting by Stochastic Transmit Beamforming and Alamouti Space-Time Coding''} \\
~ \\
\singlespacing
Technical Report \\
Department of Electronic Engineering, \\
The Chinese University of Hong Kong,
Hong Kong \\
~ \\
Sissi Xiaoxiao Wu$^\dag$, Wing-Kin Ma$^\dag$, and Anthony Man-Cho So$^\S$ \\
~ \\
\today
~ \\ ~ \\
{\footnotesize
\begin{tabular}[t]{l}
$^\dag$Department of Electronic Engineering, The Chinese University of Hong Kong, Shatin, Hong Kong. \\
E-mail: xxwu@ee.cuhk.edu.hk, wkma@ieee.org. \\
$^\S$Department of Systems Engineering and Engineering Management, The Chinese University of Hong Kong, \\
Shatin, Hong Kong. \\
E-mail: manchoso@se.cuhk.edu.hk.
\end{tabular}}
\end{center}

\vspace*{0.25in}
\noindent {\bf Abstract---}
This is a companion technical report of the main manuscript
``Physical-Layer Multicasting by Stochastic Transmit Beamforming and Alamouti Space-Time Coding''.
The report serves to give detailed derivations of the achievable rate functions encountered in the main manuscript, which are too long to be included in the latter.
In addition, more simulation results are presented to verify the viability of the multicast schemes developed in the main manuscript.
\vspace*{0.25in}

\newpage

In the main paper~\cite{MainPaper}, we establish several efficient and provably good strategies for physical-layer multicasting in multi-input single-output downlink.
In one strategy, called stochastic beamforming (SBF),
we characterize the performance of several SBF schemes by analyzing their multicast achievable rates.
In the main paper, we place emphasis on extracting insight from the SBF achievable rate expressions, and omit the details of the derivations.
The role of this technical report is to provide the detailed solutions to those achievable rate expressions.
Also, we take this opportunity to show more simulation results to further support the viability of the proposed multicast schemes.

\section{Achievable Rate Gap Analysis of SBF Schemes}

\subsection{The Gaussian SBF Scheme}
Recall from~\cite[Section~III.C]{MainPaper} that the achievable rate gap of the Gaussian SBF scheme is given by
\begin{eqnarray}
g_{\sf SBF}^{\sf Gauss}(P) &=& C_{\sf MC}(P) - C_{\sf SBF}^{\sf Gauss}(P) \nonumber \\
\noalign{\medskip}
&=& \log( 1 + \rho_{\rm min} P ) - e^{1/(\rho_{\rm min} P)} E_1( 1/(\rho_{\rm min} P) ),  \label{eq:SBF_Gaus}
\end{eqnarray}
where $E_1(x) = \int_1^\infty t^{-1} e^{-xt} dt$, $x \geq 0$, is the exponential integral of the first order.  It is known that
\begin{equation} \label{eq:E1}
E_1(x) = -\gamma - \log(x) - \sum_{k=1}^\infty \frac{(-x)^k}{k \cdot k! },
\end{equation}
where $\gamma= -\int_0^\infty \log(x) e^{-x} dx = 0.5772$ is the Euler constant; see~\cite[p.~229]{abramowitz+stegun}.  By substituting \eqref{eq:E1} into \eqref{eq:SBF_Gaus} and using the identity $e^{1/\beta} = \sum_{k=0}^\infty 1/(\beta^k k! )$, we obtain
\begin{equation} \label{eq:g_beta_exp}
g_{\sf SBF}^{\sf Gauss}(P) = \log(1+ \rho_{\rm min} P) - \log( \rho_{\rm min} P) + e^{1/(\rho_{\rm min} P)} \gamma + h(P),
\end{equation}
where
\[ h(P) = - \sum_{k=1}^\infty \frac{\log(\rho_{\rm min} P)}{(\rho_{\rm min} P)^k \cdot k!} + \sum_{k=1}^\infty \frac{e^{1/(\rho_{\rm min} P)}}{ (-\rho_{\rm min} P)^k \cdot k \cdot k! }. \]
Since $\lim_{\beta \rightarrow \infty} \log(\beta)/\beta^k = 0$ for any $k \geq 1$, one can readily verify that $\lim_{P \rightarrow \infty} h(P) = 0$.
Consequently, we have from \eqref{eq:g_beta_exp} that
\[ \lim_{P \rightarrow \infty} g_{\sf SBF}^{\sf Gauss}(P) = \gamma. \]

\subsection{The Elliptic SBF Scheme}
\label{sec:ellip-rate}
In~\cite[Section III.D]{MainPaper}, we showed that the achievable rate function for the elliptic SBF scheme is given by
\begin{equation} \label{eq:ellip_rate_raw}
C_{\sf SBF}^{\sf Ellip}(P) = \left( 1 - \frac{1}{r} \right) \int_0^r \log( 1 + t \rho_{\rm min} P ) \left( 1 - \frac{t}{r} \right)^{r-2} dt.
\end{equation}
\subsubsection{Proof of \cite[Proposition 2]{MainPaper}}
Our goal now is to show that for any $P>0$,
$$ %\begin{equation} \label{eq:ellip_rate}
C_{\sf SBF}^{\sf Ellip}(P) = \left( 1 + \frac{1}{r\rho_{\rm min}P} \right)^{r-1} \left[ \log(1+r\rho_{\rm min}P) - \sum_{k=1}^{r-1}\frac{1}{k} - \sum_{k=1}^{r-1} {r-1 \choose k} \frac{(-1)^k}{k(1+r\rho_{\rm min}P)^k} \right].
$$ %\end{equation}
Towards that end, we use \eqref{eq:ellip_rate_raw} to compute
\begin{align}
   C_{\sf SBF}^{\sf Ellip}(P) &= \int_0^r \frac{\rho_{\rm min} P}{1+t\rho_{\rm min}P} \left( 1-\frac{t}{r} \right)^{r-1}dt \label{eq:ellip-ibp} \\
   \noalign{\medskip}
   &= \int_1^{1+r\rho_{\rm min} P} \frac{1}{y} \left( 1 - \frac{y-1}{r\rho_{\rm min} P} \right)^{r-1} dy \label{eq:ellip-cvar} \\
   \noalign{\medskip}
   &= \left( 1 + \frac{1}{r\rho_{\rm min}P} \right)^{r-1} \int_1^{1+r\rho_{\rm min}P} \frac{1}{y} \left( 1 - \frac{y}{1+r\rho_{\rm min}P} \right)^{r-1} dy \nonumber \\
   \noalign{\medskip}
   &= \left( 1 + \frac{1}{r\rho_{\rm min}P} \right)^{r-1} \int_1^{1+r\rho_{\rm min}P} \left[ \frac{1}{y} + \sum_{k=1}^{r-1} {r-1 \choose k} (-1)^k \frac{y^{k-1}}{(1+r\rho_{\rm min}P)^k} \right] dy \label{eq:ellip-bin-thm} \\
   \noalign{\medskip}
   &= \left( 1 + \frac{1}{r\rho_{\rm min}P} \right)^{r-1} \left[ \log(1+r\rho_{\rm min}P) + \sum_{k=1}^{r-1} {r-1 \choose k} \frac{(-1)^k}{k} \left( 1 - \frac{1}{(1+r\rho_{\rm min}P)^k} \right) \right] \nonumber \\
   \noalign{\medskip}
   &= \left( 1 + \frac{1}{r\rho_{\rm min}P} \right)^{r-1} \left[ \log(1+r\rho_{\rm min}P) - \sum_{k=1}^{r-1}\frac{1}{k} - \sum_{k=1}^{r-1} {r-1 \choose k} \frac{(-1)^k}{k(1+r\rho_{\rm min}P)^k} \right], \label{eq:ellip-har-sum}
\end{align}
where \eqref{eq:ellip-ibp} follows from integration by parts; \eqref{eq:ellip-cvar} follows from the change of variable $y=1+t\rho_{\rm min}P$; \eqref{eq:ellip-bin-thm} follows from the binomial theorem; and \eqref{eq:ellip-har-sum} follows from the identity
\begin{equation} \label{eq:bin-sum-1}
   \sum_{k=1}^n {n \choose k} \frac{(-1)^k}{k} = -\sum_{k=1}^n \frac{1}{k};
\end{equation}
see~\cite[Formula 0.155(4)]{bk:Gradshteyn}.  This completes the proof.

\subsubsection{Derivation of $\lim_{P \rightarrow \infty} g_{\sf SBF}^{\sf Ellip}(P)$}
\medskip
\noindent The achievable rate gap of the elliptic SBF scheme is given by
\begin{eqnarray*}
g_{\sf SBF}^{\sf Ellip}(P) &=& C_{\sf MC}(P) - C_{\sf SBF}^{\sf Ellip}(P) \\% \nonumber \\
\noalign{\medskip}
&=& \log(1+\rho_{\rm min}P) \\ %\nonumber \\
\noalign{\medskip}
&\quad-& \left( 1 + \frac{1}{r\rho_{\rm min}P} \right)^{r-1} \left[ \log(1+r\rho_{\rm min}P) - \sum_{k=1}^{r-1}\frac{1}{k} - \sum_{k=1}^{r-1} {r-1 \choose k} \frac{(-1)^k}{k(1+r\rho_{\rm min}P)^k} \right].  % \label{eq:ellip_rate_gap}
\end{eqnarray*}
We claim that
$$ \lim_{P \rightarrow \infty} g_{\sf SBF}^{\sf Ellip}(P) = \sum_{k=1}^{r-1} \frac{1}{k} - \log(r). $$
This essentially follows by applying the l'H\^{o}pital rule.  Alternatively, from \eqref{eq:ellip-ibp}, we have
\begin{align*}
   C_{\sf SBF}^{\sf Ellip}(P) &= \int_0^r \frac{\rho_{\rm min} P}{1+t\rho_{\rm min}P} \left( 1-\frac{t}{r} \right)^{r-1}dt \\
   \noalign{\medskip}
   &=\log(1+r\rho_{\rm min}P) + \sum_{k=1}^{r-1} {r-1 \choose k} \frac{(-1)^k}{r^k} \int_0^r \frac{t^k\rho_{\rm min} P}{1+t\rho_{\rm min}P} dt
\end{align*}
by the binomial theorem.  Hence, by the dominated convergence theorem and \eqref{eq:bin-sum-1}, we obtain
\begin{align*}
   \lim_{P \rightarrow \infty} g_{\sf SBF}^{\sf Ellip}(P) &= \lim_{P \rightarrow \infty} \left[ \log(1+\rho_{\rm min}P) - \log(1+r\rho_{\rm min}P) - \sum_{k=1}^{r-1} {r-1 \choose k} \frac{(-1)^k}{r^k} \int_0^r \frac{t^k\rho_{\rm min} P}{1+t\rho_{\rm min}P} dt \right] \\
   \noalign{\medskip}
   &= -\log(r) - \sum_{k=1}^{r-1} {r-1 \choose k} \frac{(-1)^k}{r^k} \int_0^r t^{k-1} dt \\
   \noalign{\medskip}
   &= -\log(r) + \sum_{k=1}^{r-1} \frac{1}{k},
\end{align*}
as required.

\section{Achievable Rate Analysis of the Bingham SBF Scheme (Proof of~\cite[Proposition 4]{MainPaper})}

For the Bingham SBF scheme, we showed in~\cite[Section~III.E, Proposition~3]{MainPaper} that the rate of user~$i$ is given by
$$ %\begin{equation} \label{eq:bing_rate}
C_{{\sf SBF},i}^{\sf Bing}(P) = \mathbb{E}_{\bm{w}}[ \log( 1 + P | {\bf h}_i^H {\bm w} |^2 ) ]
    = \log( 1 + \rho_i P ) +
    %\varphi\left( \frac{\bm{\mu}_i}{\sum_{k=1}^r \mu_{i,k}} \right)
    \varphi\left( \frac{\bm{\mu}_i}{ \bm{\mu}_i^T {\bf 1} } \right)
    - \varphi\left( \bm{\lambda} \right),
$$ %\end{equation}
where $\varphi: \mathbb{R}^r \rightarrow \mathbb{R}$ is given by
$$ %\begin{equation} \label{eq:varphi}
\varphi\left( {\bf d} \right) = \mathbb{E}_{\bm{\zeta}}\left[ \log\left( \sum_{k=1}^r d_k \zeta_k \right)   \right],
$$ %\end{equation}
and $\bm{\zeta}$ is a random vector with independent and identical unit-mean exponentially distributed components.  Hence, in order to derive an explicit expression for $C_{{\sf SBF},i}^{\sf Bing}(P)$, it suffices to derive one for $\varphi$.  Towards that end, let $\zeta=\sum_{k=1}^r d_k \zeta_k$.  We can then express $\varphi$ as
\begin{equation} \label{eq:t_sumofexp_2}
\varphi\left( {\bf d} \right) = \int_{0}^{\infty} \log(z) p_\zeta(z) dz.
\end{equation}
According to \cite{Bjornson2010,Jnl:sumofexp}, $\zeta$ follows the distribution
\begin{equation} \label{eq:t_sumofexp_3}
p_\zeta (z) =
\prod_{n=1}^{c}\frac{1}{\tilde d_n^{r_n}}
\sum_{k=1}^{c}\sum_{m=1}^{r_k}\frac{\Psi_{k,m,\mathbf{r}}}{(r_k-m)!}(-1)^{(r_k-m)}z^{(r_k-m)}e^{-z/\tilde d_k}, \quad z \geq 0,
\end{equation}
where $\Psi_{k,m,\mathbf{r}}$ is defined in~\cite[Proposition 4]{MainPaper}.
Substituting \eqref{eq:t_sumofexp_3} into \eqref{eq:t_sumofexp_2} yields
\begin{equation}
\label{eq:t_sumofexp_0}
\varphi\left( {\bf d} \right) =
\prod_{n=1}^{c}\frac{1}{\tilde d_n^{r_n}}
\sum_{k=1}^{c}\sum_{m=1}^{r_k}\frac{\Psi_{k,m,\mathbf{r}}}{(r_k-m)!}(-1)^{(r_k-m)}
\int_{0}^{\infty} z^{(r_k-m)} e^{-z/\tilde d_k} \log z \, dz.
\end{equation}
By the result in \cite[Formula 4.352(2)]{bk:Gradshteyn}, we have
\begin{equation} \label{eq:t_sumofexp}
\int_{0}^{\infty}z^{(r_k-m)}e^{-z/\tilde d_k}\log z \, dz = \theta({\tilde d}_k, r_k-m),
\end{equation}
where $\theta(\cdot,\cdot)$ is defined in~\cite[Proposition 4]{MainPaper}.  Putting \eqref{eq:t_sumofexp} in \eqref{eq:t_sumofexp_0}, the proof is completed.
\vspace{\baselineskip}

%While Proposition~\ref{fact:sumofexp} gives an explicit expression of \eqref{eq:varphi}, which in turn provides a way of computing the Bingham SBF achievable rate efficiently (in comparison to using Monte Carlo simulations), it is too complicated to extract insight. This difficulty motivated us to turn to the stochastic majorization technique for Bingham SBF rate gap characterization \cite[Section~III.E]{MainPaper}.

\section{Achievable Rate Gap Analysis of SBF Alamouti Schemes}
\subsection{The Gaussian SBF Alamouti Scheme (Proof of~\cite[Proposition 6]{MainPaper})}

Recall from~\cite[Section V.B]{MainPaper} that the achievable rate of the Gaussian SBF Alamouti scheme is given by
$$ C_{\sf SBF-ALAM}^{\sf Gauss}(P) = \mathbb{E}_{\xi }[\log (1 + \xi\rho_{\rm min} P )], $$
where $\xi$ follows a chi-square distribution with unit mean and 4 degrees of freedom.  Since the PDF of $\xi$ is
$$ p_{\xi}(t) = 4t e^{-2t} \quad\mbox{for } t \geq 0, $$
we have
$$
 C_{\sf SBF-ALAM}^{\sf Gauss}(P) = 4 \int_0^\infty \log(1+t\rho_{\rm min}P) te^{-2t}\,dt = \frac{4}{(\rho_{\rm min}P)^2} \int_0^{\infty} \log(1+t)te^{-\tfrac{2t}{\rho_{\rm min}P}}\,dt.
$$
By the result in~\cite[Appendix B]{Alouini1999},
$$ \int_0^{\infty} \log(1+t)te^{-\tfrac{2t}{\rho_{\rm min}P}}\,dt = e^{\tfrac{2}{\rho_{\rm min}P}} \left[ \frac{\Gamma(-1,2/(\rho_{\rm min}P))}{2/(\rho_{\rm min}P)} + \frac{\Gamma(0,2/(\rho_{\rm min}P))}{(2/(\rho_{\rm min}P))^2}\right], $$
where $\Gamma(\alpha,x) = \int_x^\infty t^{\alpha-1} e^{-t} dt$ is the complementary incomplete gamma function.  It follows that
\begin{equation} \label{eq:Gauss_SBF_alam_0}
C_{\sf SBF-ALAM}^{\sf Gauss}(P) = e^{\tfrac{2}{\rho_{\rm min} P}}
\left[ \Gamma(0,2/(\rho_{\rm min} P))  +
\frac{ \Gamma(-1,2/(\rho_{\rm min} P)) }{ (\rho_{\rm min} P)/2 }
\right].
\end{equation}
Using the identities $\Gamma(0,x) = E_1(x)$ and $\Gamma(0,x)= -\Gamma(-1,x) + x^{-1} e^{-x}$,
we can simplify \eqref{eq:Gauss_SBF_alam_0} to
$$ %\begin{equation} \label{eq:Gauss_SBF_alam}
C_{\sf SBF-ALAM}^{\sf Gauss}(P)=  \left( 1 - \tfrac{2}{\rho_{\rm min} P} \right) e^{\tfrac{2}{\rho_{\rm min} P}} E_1 \left(\tfrac{2}{\rho_{\rm min} P}\right) + 1.
$$ %\end{equation}
Now, let us examine the limit of $g_{\sf SBF-ALAM}^{\sf Gauss}(P) = C_{\sf MC}(P)- C_{\sf SBF-ALAM}^{\sf Gauss}(P)$ as $P \rightarrow \infty$.
Following the idea in the derivation of \eqref{eq:g_beta_exp}, we write
\begin{equation} \label{eq:g_Gauss_SBF_alam}
g_{\sf SBF-ALAM}^{\sf Gauss}(P)=
    \log( 1 + \rho_{\min} P )
    - \left( 1 - \frac{2}{\rho_{\rm min} P} \right) \log\left( \frac{\rho_{\rm min} P}{2} \right)
    + \left( 1 - \frac{2}{\rho_{\rm min} P} \right) e^{\tfrac{2}{\rho_{\rm min} P}} \gamma
    + h(P) - 1,
\end{equation}
where
\[ h(P) = \left( 1 - \frac{2}{\rho_{\rm min} P} \right) \left( - \sum_{k=1}^\infty \frac{\log( \rho_{\rm min} P/2 )}{(\rho_{\rm min} P/2)^k \cdot k!} + \sum_{k=1}^\infty \frac{(-1)^k \cdot e^{2/(\rho_{\rm min} P)}}{ (\rho_{\rm min} P/2)^k \cdot k \cdot k! } \right). \]
One can show that $\lim_{P \rightarrow \infty} h(P) = 0$.
As a result, we deduce from \eqref{eq:g_Gauss_SBF_alam} that
\[ \lim_{P \rightarrow \infty} g_{\sf SBF-ALAM}^{\sf Gauss}(P) = \log(2) + \gamma - 1.
\]

\subsection{The Elliptic SBF Alamouti Scheme (Proof of~\cite[Proposition 8]{MainPaper})}
Recall from~\cite[Section V.C]{MainPaper} that the achievable rate of the elliptic SBF Alamouti scheme is given by
\begin{equation} \label{eq:ellip-alam-rate-int}
   C_{\sf SBF-ALAM}^{\sf Ellip}(P) = \frac{(2r-1)(2r-2)}{r} \int_0^r \log(1+t\rho_{\rm min}P) \cdot \frac{t}{r} \left( 1 - \frac{t}{r} \right)^{2r-3}dt.
\end{equation}
Let
\begin{eqnarray*}
   C_1(P) &=& \frac{(2r-1)(2r-2)}{r} \int_0^r \log(1+t\rho_{\rm min}P) \cdot \left( 1 - \frac{t}{r} \right)^{2r-3}dt, \\
   \noalign{\medskip}
   C_2(P) &=& \frac{(2r-1)(2r-2)}{r} \int_0^r \log(1+t\rho_{\rm min}P) \cdot \left( 1 - \frac{t}{r} \right)^{2r-2}dt.
\end{eqnarray*}
Then, we have
$$  C_{\sf SBF-ALAM}^{\sf Ellip}(P) = C_1(P) - C_2(P). $$
Now, following the techniques in Section \ref{sec:ellip-rate}, we compute
\begin{align*}
   C_1(P) &= (2r-1) \int_0^r \frac{\rho_{\rm min}P}{1+t\rho_{\rm min}P} \left( 1 - \frac{t}{r} \right)^{2r-2}dt \\
   \noalign{\medskip}
   &= (2r-1)\left( 1+\frac{1}{r\rho_{\rm min}P} \right)^{2r-2} \int_1^{1+r\rho_{\rm min}P} \frac{1}{y} \left( 1 - \frac{y}{1+r\rho_{\rm min}P} \right)^{2r-2}dy \\
   \noalign{\medskip}
   &= (2r-1)\left( 1+\frac{1}{r\rho_{\rm min}P} \right)^{2r-2} \left[ \log(1+r\rho_{\rm min}P) - \sum_{k=1}^{2r-2} \frac{1}{k} - \sum_{k=1}^{2r-2} {2r-2 \choose k} \frac{(-1)^k}{k(1+r\rho_{\rm min}P)^k} \right]
\end{align*}
and
\begin{align*}
   C_2(P) &= (2r-2) \int_0^r \frac{\rho_{\rm min}P}{1+t\rho_{\rm min}P} \left( 1 - \frac{t}{r} \right)^{2r-1}dt \\
   \noalign{\medskip}
   &= (2r-2) \left( 1 + \frac{1}{r\rho_{\rm min}P} \right)^{2r-1}  \int_1^{1+r\rho_{\rm min}P} \frac{1}{y} \left( 1 - \frac{y}{1+r\rho_{\rm min}P} \right)^{2r-1}dy \\
   \noalign{\medskip}
   &= (2r-2)\left( 1+\frac{1}{r\rho_{\rm min}P} \right)^{2r-1} \left[ \log(1+r\rho_{\rm min}P) - \sum_{k=1}^{2r-1} \frac{1}{k} - \sum_{k=1}^{2r-1} {2r-1 \choose k} \frac{(-1)^k}{k(1+r\rho_{\rm min}P)^k} \right].
\end{align*}
Thus, we have arrived at the following explicit formula for $C_{\sf SBF-ALAM}^{\sf Ellip}(P)$:
\begin{align}
   & C_{\sf SBF-ALAM}^{\sf Ellip}(P) \nonumber \\
   \noalign{\medskip}
   &= (2r-1)\left( 1+\frac{1}{r\rho_{\rm min}P} \right)^{2r-2} \left[ \log(1+r\rho_{\rm min}P) - \sum_{k=1}^{2r-2} \frac{1}{k} - \sum_{k=1}^{2r-2} {2r-2 \choose k} \frac{(-1)^k}{k(1+r\rho_{\rm min}P)^k} \right] \nonumber \\
   \noalign{\medskip}
   &\quad- (2r-2)\left( 1+\frac{1}{r\rho_{\rm min}P} \right)^{2r-1} \left[ \log(1+r\rho_{\rm min}P) - \sum_{k=1}^{2r-1} \frac{1}{k} - \sum_{k=1}^{2r-1} {2r-1 \choose k} \frac{(-1)^k}{k(1+r\rho_{\rm min}P)^k} \right]. \label{eq:ellip_alam_rate}
\end{align}
Next, we show that the achievable rate gap $g_{\sf SBF-ALAM}^{\sf Ellip}(P) = C_{\sf MC}(P) - C_{\sf SBF-ALAM}^{\sf Ellip}(P)$ of the elliptic SBF Alamouti scheme satisfies
$$ %\begin{equation} \label{eq:ellip_alam_lim}
\lim_{P \rightarrow \infty} g_{\sf SBF-ALAM}^{\sf Ellip}(P) = \sum_{k=1}^{2r-1} \frac{1}{k} - \log(r) - 1.
$$ % end{equation}
Upon recalling that $C_{\sf MC}(P) = \log( 1+\rho_{\rm min}P )$, the desired result can be obtained by applying the l'H\^{o}pital rule to \eqref{eq:ellip_alam_rate}.  Alternatively, we can proceed as follows.  First, Proposition 4 of~\cite{MainPaper} implies that
$$ \frac{(2r-1)(2r-2)}{r} \int_0^r \frac{t}{r} \left( 1 - \frac{t}{r} \right)^{2r-3}dt = 1. $$
Hence, using the integral representation \eqref{eq:ellip-alam-rate-int}, we have
$$ g_{\sf SBF-ALAM}^{\sf Ellip}(P) = \frac{(2r-1)(2r-2)}{r} \int_0^r \log\left( \frac{1+\rho_{\rm min}P}{1+t\rho_{\rm min}P} \right) \cdot \frac{t}{r} \left( 1 - \frac{t}{r} \right)^{2r-3}dt. $$
Now, by the dominated convergence theorem and the binomial theorem, we have
\begin{align*}
   \lim_{P \rightarrow \infty} g_{\sf SBF-ALAM}^{\sf Ellip}(P) &= -\frac{(2r-1)(2r-2)}{r} \int_0^r \frac{t\log(t)}{r} \left( 1 - \frac{t}{r} \right)^{2r-3}dt \\
   \noalign{\medskip}
   &= -\frac{(2r-1)(2r-2)}{r^2}\sum_{k=0}^{2r-3} {{2r-3}\choose k} \frac{(-1)^k}{r^k}\int_0^r t^{k+1}\log(t) \, dt \\
   \noalign{\medskip}
   &= -(2r-1)(2r-2)\sum_{k=0}^{2r-3} {{2r-3}\choose k} (-1)^k \left[ \frac{\log (r)}{k+2} - \frac{1}{(k+2)^2} \right].
\end{align*}
The last sum can be computed using the following result:
\begin{Prop} \label{prop:bin-id}
   The following identities hold:
\begin{eqnarray}
   \sum_{k=0}^n {n \choose k} \frac{(-1)^k}{k+2} &=& \frac{1}{(n+2)(n+1)}, \label{eq:bin-id-1} \\
   \noalign{\medskip}
   \sum_{k=0}^n {n \choose k} \frac{(-1)^k}{(k+2)^2} &=& \frac{1}{(n+2)(n+1)} \left( \sum_{k=1}^{n+2} \frac{1}{k} - 1 \right). \label{eq:bin-id-2}
\end{eqnarray}
\end{Prop}
Assuming Proposition \ref{prop:bin-id}, we can immediately conclude that
$$ \lim_{P \rightarrow \infty} g_{\sf SBF-ALAM}^{\sf Ellip}(P) = \sum_{k=1}^{2r-1} \frac{1}{k} - \log(r) - 1, $$
as required.  Thus, it remains to prove Proposition \ref{prop:bin-id}.

\medskip
\noindent {\it Proof of Proposition \ref{prop:bin-id}:} We begin with \eqref{eq:bin-id-1}.  Using the binomial theorem, we have
\begin{eqnarray*}
   \sum_{k=0}^n {n \choose k} \frac{(-1)^k}{k+2} &=& \sum_{k=0}^n {n \choose k} (-1)^k \left( \int_0^1 x^{k+1}dx \right) \\
   \noalign{\medskip}
   &=& \int_0^1 x \left[ \sum_{k=0}^n {n \choose k} (-1)^k x^k \right] dx \\
   \noalign{\medskip}
   &=& \int_0^1 x(1-x)^n dx \\
   \noalign{\medskip}
   &=& \frac{1}{(n+2)(n+1)}.
\end{eqnarray*}
Next, we prove \eqref{eq:bin-id-2}.  Using similar techniques as above, we compute
\begin{align}
   \sum_{k=0}^n {n \choose k} \frac{(-1)^k}{(k+2)^2} &= \int_0^1\int_0^1 xy(1-xy)^n\,dxdy \nonumber \\
   \noalign{\medskip}
   &= -\frac{1}{n+1}\int_0^1 (1-y)^{n+1}dy + \frac{1}{(n+2)(n+1)}\int_0^1 \frac{1-(1-y)^{n+2}}{y}dy \label{eq:bin-id-2-ibp} \\
   \noalign{\medskip}
   &= -\frac{1}{(n+2)(n+1)} \left[ \sum_{k=1}^{n+2} {n \choose k} \frac{(-1)^k}{k} + 1 \right]  \label{eq:bin-id-2-bin} \\
   \noalign{\medskip}
   &= \frac{1}{(n+2)(n+1)} \left( \sum_{k=1}^{n+2} \frac{1}{k} - 1 \right), \label{eq:bin-id-2-har-id}
\end{align}
where \eqref{eq:bin-id-2-ibp} follows from integration by parts; \eqref{eq:bin-id-2-bin} follows from the binomial theorem; and \eqref{eq:bin-id-2-har-id} follows from the identity \eqref{eq:bin-sum-1}.  This completes the proof.
\hfill $\blacksquare$

\section{Further Simulation Results}

\subsection{Further Simulation Results for the BERs in Section VI}
We continue the
% BER
simulation comparisons in Section VI of the main paper~\cite{MainPaper}.
%by trying a few more system settings.
Our purpose is to give more empirical evidence to support the viability of the proposed multicast schemes.

\begin{figure}[htp]
\centering
\begin{center}
\includegraphics[width = 0.7\textwidth]{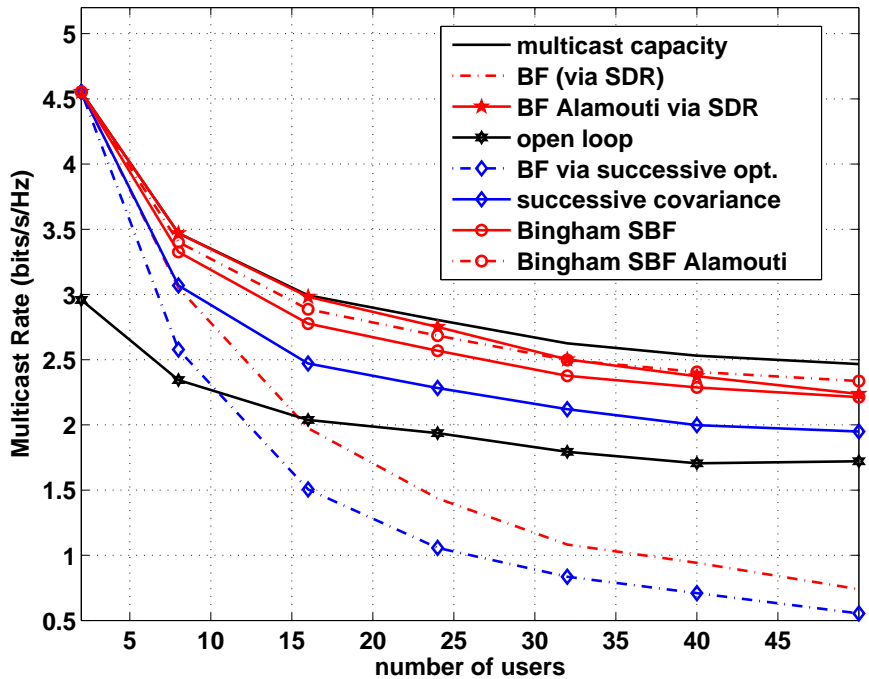}
\end{center}
%\vspace*{-\baselineskip}
\caption{Comparison with the successive optimization algorithms.}
\label{fig:successive_optimization}
\end{figure}

We compare the multicast achievable rates of our proposed schemes and the successive methods in \cite{Jnl:successive_BF_2011}.
We set $N=4$, $P= 10$dB.
The other simulation settings are identical to that in Section VI.A of \cite{MainPaper}.
Fig.~\ref{fig:successive_optimization} plots the multicast rates of the various schemes against the number of users $M$.
In the figure,
``BF via successive opt.'' and  ``successive covariance'' stand for the successive beamformer and covariance optimization designs in \cite{Jnl:successive_BF_2011}, respectively.
We can see that the proposed schemes have better multicast achievable rates than the two successive methods.

\begin{figure}[htp]
\centering
\begin{center}
\subfigure[][$M= 16$, QPSK]{\resizebox{0.485\textwidth}{!}{\includegraphics{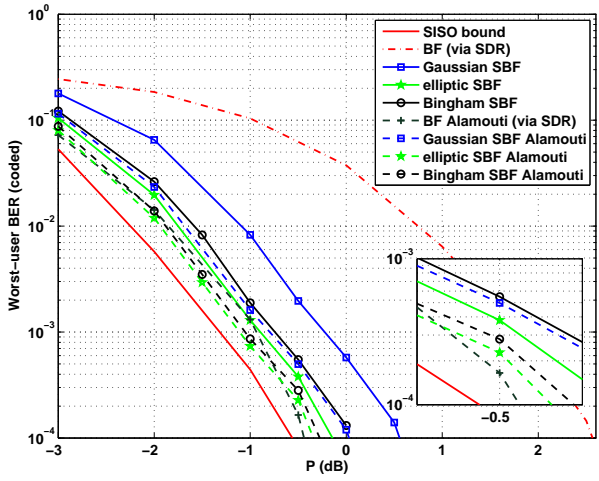}}}
\subfigure[][$M= 16$, $16$-ary QAM]{\resizebox{0.485\textwidth}{!}{\includegraphics{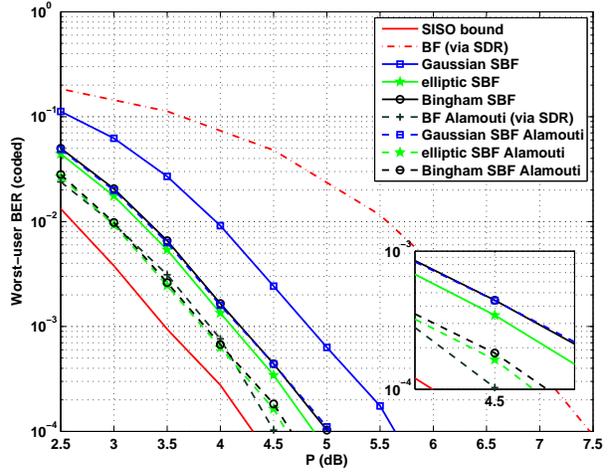}}}
\subfigure[][$M= 24$, QPSK]{\resizebox{0.485\textwidth}{!}{\includegraphics{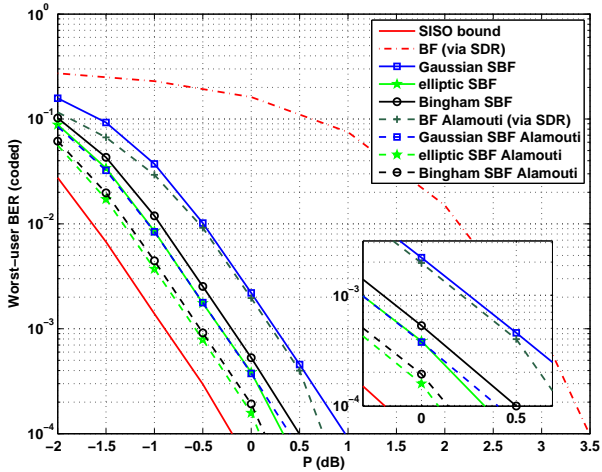}}}
\subfigure[][$M= 24$, $16$-ary QAM]{\resizebox{0.485\textwidth}{!}{\includegraphics{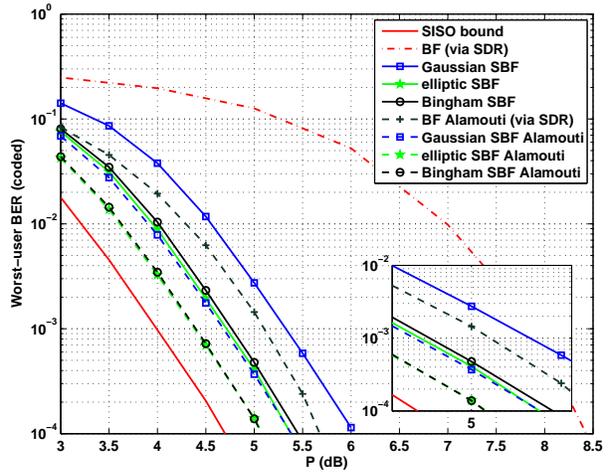}}}
\end{center}
\caption{The worst-user BER performance of the various multicast schemes.}
\label{fig:BER}
\end{figure}

%, by trying a few more different settings.
Next, we provide more BER simulations.
The simulation settings are the same as those in Fig. 2 of \cite{MainPaper}.
The results are given in Fig.~\ref{fig:BER},
where we try two different values of the number of users $M$ and two types of symbol constellations.
The two values of $M$ are $M= 16$ and $M= 24$; the two symbol constellations are QPSK and $16$-ary QAM.
Note that the frame lengths are $T= 1440$ for QPSK, and $T= 720$ for $16$-ary QAM.
From Fig.~\ref{fig:BER}, we observe that the BER performance of the various multicast schemes under $16$-ary QAM follows almost the same trend as that under QPSK.
The other observations are analogous to what we have reported in the main paper:
The beamformed Alamouti scheme has an edge when there is a smaller number of users (under both QPSK and $16$-ary QAM).
%The number of users has little impact on the performance of the SBF schemes and the SBF Alamouti schemes.
The performance of the SBF schemes and the SBF Alamouti schemes is relatively insensitive to the number of users.
The elliptic SBF Alamouti scheme generally exhibits the best performance.

\subsection{Further Simulation Results: Precoded Spatial Multiplexing and Precoded Quasi-Orthogonal Space-Time Block Coding}

The aim of this section is to implement some intuition-driven ideas that were previously untested in the literature (to the best of our knowledge),
and examine how they work in terms of BER performance.
Specifically, we consider combined schemes where spatial multiplexing (SM) or a ``good'' isotropic space-time code is precoded (or transmit-beamformed) such that the corresponding transmit signal covariance equals the multicast capacity-optimal transmit covariance.
At first, one would be tempted to think that such schemes should yield good BER performance.
To this end, we will show by simulations that this is not the case.

We consider two schemes.
The first is a precoded SM scheme in which the transmit signal is given by
\begin{equation} \label{eq:xt_SM}
{\bf x}(t) = \sqrt{P} {\bf B} {\bf s}(t),
\end{equation}
where ${\bf B} \in \mathbb{C}^{N \times d}$ is the precoding matrix,
${\bf s}(t) = ( s_1(t),\ldots,s_d(t)) \in \mathbb{C}^d$ is a vector containing multiple data streams,
and $d$ is the number of streams.
We consider a precoding matrix that is optimal from a multicast capacity perspective:
Choose ${\bf B}$ as a square root factor of the multicast capacity-optimal transmit covariance; i.e.,
\[ {\bf B}{\bf B}^H = {\bf W}^\star, \]
where ${\bf W}^\star$ is the multicast capacity-optimal transmit covariance, determined by solving Problem (MC) in \cite{MainPaper}
and $d= {\rm rank}({\bf W}^\star)$.
Note that from an achievable rate viewpoint,
such precoded SM scheme is multicast capacity-optimal under Gaussian ${\bf s}(t)$.
However, %for coded BER simulations,
in practice,
${\bf s}(t)$ is often
%generated from coded modulation schemes,
drawn from finite symbol constellations.
We employ bit interleaved coded modulation (BICM) to generate the symbol vector ${\bf s}(t)$,
using the same way as in the widely adopted BICM SM scheme for point-to-point MIMO~\cite{Makay_Collings_05}.
The maximum-likelihood (ML) detector is used to detect ${\bf s}(t)$, and in our simulations we use the (optimal) exhaustive search to implement the ML detector.

{\sloppy
The second scheme combines precoding and the quasi-orthogonal space-time block code (QOSTBC)~\cite{QOSTBC_Jafar2001}.
For ease of exposition, consider that the number of transmit antennas is $N= 4$ (extension to $N=8$ is possible).
Given a data symbol sequence $\{ s(t) \}_{t=1}^T$,
we parse it into blocks via $\bm{s}(n) = [~ s(4n), s(4n+1), s(4n+2), s(4n+3) ~]^T$.
Likewise, we parse the transmit signal into blocks:
\[
{\bm X}(n) = [~ {\bf x}(4n), ~ {\bf x}(4n+1), ~{\bf x}(4n+2), ~{\bf x}(4n+3) ~].
\]
For each transmit code block ${\bm X}(n)$, we apply a beamformed QOSTBC}
\[
{\bm X}(n) = \sqrt{P} {\bf B} {\bf C}( \bm{s}(n) ),
\]
where ${\bf B}$ is the precoding matrix, chosen to satisfy the multicast capacity-optimal transmit covariance ${\bf B}{\bf B}^H = {\bf W}^\star$,
and
\[
{\bf C}({\bm s}) = \begin{bmatrix} s_1 & s_2 & s_3 & s_4 \\ -s_2^* & s_1^* & -s_4^* & s_3^* \\ -s_3^* & -s_4^* & s_1^* & s_2^*\\ s_4 & -s_3 & -s_2 & ~s_1\end{bmatrix}
\]
is the $4 \times 4$ QOSTBC~\cite{QOSTBC_Jafar2001}.
Like the precoded SM scheme, this precoded QOSTBC scheme seems to be good in the sense that it satisfies the multicast capacity-optimal transmit covariance property $\frac{1}{4} \mathbb{E}[ {\bm X}(n) {\bm X}^H(n) ] =P {\bf B}{\bf B}^H = P {\bf W}^\star$.
Note that precoded QOSTBC also requires ML detection at the receivers,
since QOSTBCs are not exactly orthogonal.

Next, we describe how the simulations are prepared for the precoded SM and precoded QOSTBC schemes.
In addition to the two schemes above,
the other schemes under test are the beamforming scheme and all our proposed schemes.
In order to make the comparison fair,
the bit rate should be fixed for all the schemes.
This is not a problem for the precoded QOSTBC scheme, the beamforming scheme and all our schemes.
%, since they can all be regarded as single-stream based.
However, there is an issue with the precoded SM scheme, where its number of streams depends on $d= {\rm rank}({\bf W}^\star)$.
This means that if a fixed bit rate is desired, then the symbol constellation and/or the coding rate must be changed in accordance with ${\rm rank}({\bf W}^\star)$, which can complicate the matter.
Therefore, to simplify the comparison, we evaluate the BERs for instances where  ${\rm rank}({\bf W}^\star)= 4$;
i.e., we pick up channel realizations where  ${\rm rank}({\bf W}^\star)$ equals $4$, and use them to evaluate the BERs of the various schemes.
We make the bit rate of all the schemes identical by using $16$-ary QAM symbol constellation for all the single-stream-based schemes,
and BPSK for the precoded SM scheme (which is now fixed to have $4$ streams under the current setup).
Moreover, all the schemes employ BICM, wherein the coding scheme is the Turbo code.
The other simulation settings are the same as those in the coded BER simulations described in the main manuscript.

\begin{figure}[htp]
\centering
\begin{center}
\includegraphics[width = 0.7\textwidth]{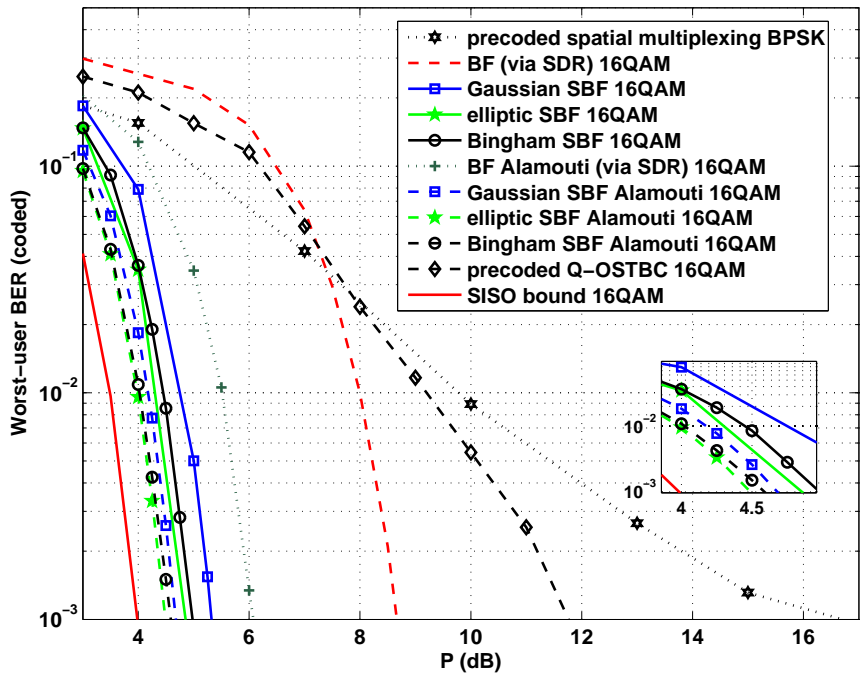}
\end{center}
%\vspace*{-\baselineskip}
\caption{Comparison with the precoded SM scheme and the precoded QOSTBC scheme.}
\label{fig:BER2}
\end{figure}

Fig.~\ref{fig:BER2} shows the results.
The number of users is $M= 32$.
We see that the precoded SM and precoded QOSTBC schemes do not work well.
In particular, for $P \geq 8$dB, even the beamforming scheme performs better than the precoded SM and QOSTBC schemes.
%At this point, we should mention that there is actually a difference between point-to-point scenarios and the considered multicast scenario---
We find that intersymbol interference (ISI) actually has much impact on the BER performance of the precoded SM and QOSTBC schemes.
In particular, it appears that some users are subject to severe ISI effects.
%, which have not been controlled in the beamformed SM and QOSTBC schemes.
By contrast, the ISI-free nature of the beamforming scheme and the proposed schemes turns out to provide performance advantages over the precoded SM and QOSTBC schemes.
%Moreover, from an information theoretic viewpoint,
%the finite constellation nature of the beamformed SM and QOSTBC schemes means that they do not truly form vector Gaussian codewords.
%In principle, this issue may be remedied by developing a more careful beamformer design in place of intuition, say, via optimization of a worst-user error probability metric.
%However, we should also stress that presently, there seems to be no such work for multicasting.

In this simulation study, we conclude that a straightforward precoded extension of SM or QOSTBC may lead to a mediocre and inefficient physical-layer multicasting scheme.
While this is our empirical observation,
it should be noted that the problem lies in the ISI, rather than the physical-layer structure itself.
In other words, it may be possible for one to improve the performance of the precoded SM or QOSTBC schemes by a more careful (and likely more sophisticated) precoder design, in which the ISI or some error probability metric are explicitly dealt with.
At present, there seems to be no work in such direction.

\bibliographystyle{IEEEbib}
\bibliography{IEEEabrv,ref}

\end{document}